\documentclass[runningheads]{llncs}
\usepackage{xcolor}
\usepackage[T1]{fontenc}
\usepackage{indentfirst}
\usepackage{hyperref}
\usepackage{url}
\usepackage{bm}
\usepackage{svg}
\def\doi#1{\href{https://doi.org/\detokenize{#1}}{\url{https://doi.org/\detokenize{#1}}}}
\usepackage{graphicx}
\usepackage{listings}
\usepackage{multirow}
\usepackage{wrapfig}
\begin{document}
%
\title{U-Net vs Transformer: Is U-Net Outdated in Medical Image Registration?}

%
\author{Xi Jia\inst{1}, Joseph Bartlett\inst{1,2}, Tianyang Zhang\inst{1}, Wenqi Lu\inst{3}, \\ Zhaowen Qiu\inst{4}, and Jinming Duan\inst{1,5}}
\institute{School of Computer Science, \\University of Birmingham, Birmingham, B15 2TT, UK \and 
Department of Biomedical Engineering, \\ University of Melbourne, Melbourne, VIC 3010, Australia \and
Tissue Image Analytics Centre, Department of Computer Science, \\University of Warwick, Coventry, CV4 7AL, UK \and 
Institute of Information Computer Engineering, \\ Northeast Forestry University, Harbin, 150400, China \\ \and
Alan Turing Institute, London, NW1 2DB, UK
}
%
%
\maketitle              

\begin{abstract}
Due to their extreme long-range modeling capability, vision transformer-based networks have become increasingly popular in deformable image registration. We believe, however, that the receptive field of a 5-layer convolutional U-Net is sufficient to capture accurate deformations without needing long-range dependencies. The purpose of this study is therefore to investigate whether U-Net-based methods are outdated compared to modern transformer-based approaches when applied to medical image registration. For this, we propose a large kernel U-Net (LKU-Net) by embedding a parallel convolutional block to a vanilla U-Net in order to enhance the effective receptive field.  On the public 3D IXI brain dataset for atlas-based registration, we show that the performance of the vanilla U-Net is already comparable with that of state-of-the-art transformer-based networks (such as TransMorph), and that the proposed LKU-Net outperforms TransMorph by using only 1.12\% of its parameters and 10.8\% of its mult-adds operations. We further evaluate LKU-Net on a MICCAI Learn2Reg 2021 challenge dataset for inter-subject registration, our LKU-Net also outperforms TransMorph on this dataset and ranks first on the public leaderboard as of the submission of this work. With only modest modifications to the vanilla U-Net, we show that U-Net can outperform transformer-based architectures on inter-subject and atlas-based 3D medical image registration. Code is available at \url{https://github.com/xi-jia/LKU-Net}.

\end{abstract}

\section{Introduction}
\noindent Deformable image registration, a fundamental task in medical image analysis, aims to find an optimal deformation that maps a moving image onto a fixed image. The problem can be formulated as the minimization problem including a data fidelity term that measures the distance between the fixed and warped moving image and a regularization that penalizes non-smooth deformations.

Many iterative optimization approaches \cite{avants_ANTS,rueckert1999nonrigid,vercauteren2009diffeomorphic} have been proposed to tackle intensity-based deformable registration, and shown great registration accuracy. However, such methods suffer from slow inference speeds and manual tuning for each new image pair. Though some works, such as Nesterov accelerated ADMM \cite{thorley2021nesterov}, propose certain techniques to accelerate the computation, their speed still does not compare to approaches based on deep learning. Due to their fast inference speed and comparable accuracy with iterative methods, registration methods based on deep neural networks  \cite{balakrishnan2018unsupervised,zhang2018inverse,balakrishnan2019voxelmorph,Zhao_2019_ICCV,Mok_2020_CVPR,qiu2021learning,jia2021learning} have become a powerful benchmark for large-scale medical image registration.

Deep learning based registration methods \cite{zhang2018inverse,balakrishnan2019voxelmorph,Mok_2020_CVPR,qiu2021learning} directly take moving and fixed image pairs as input, and output corresponding estimated deformations. Most deep neural networks use a U-Net style architecture \cite{ronneberger2015u} as their backbone and only vary preprocessing steps and loss functions. Such an architecture includes a contraction path to encode the spatial information from the input image pair, and an expansion path to decode the spatial information to compute a deformation field (or stationary velocity field). Inspired by the success of the transformer architecture \cite{NIPS2017_3f5ee243}, several recent registration works \cite{zhang2021learning,chen2021vit,chen2021transmorph} have used it as the backbone to replace the standard U-Net. In this study, we investigate whether U-Net is outdated compared to modern transformer architectures, such as the new state-of-the-art TransMorph \cite{chen2021transmorph}, for image registration.

\begin{figure}[t!]
  \centering
  \includegraphics[width=0.75\linewidth]{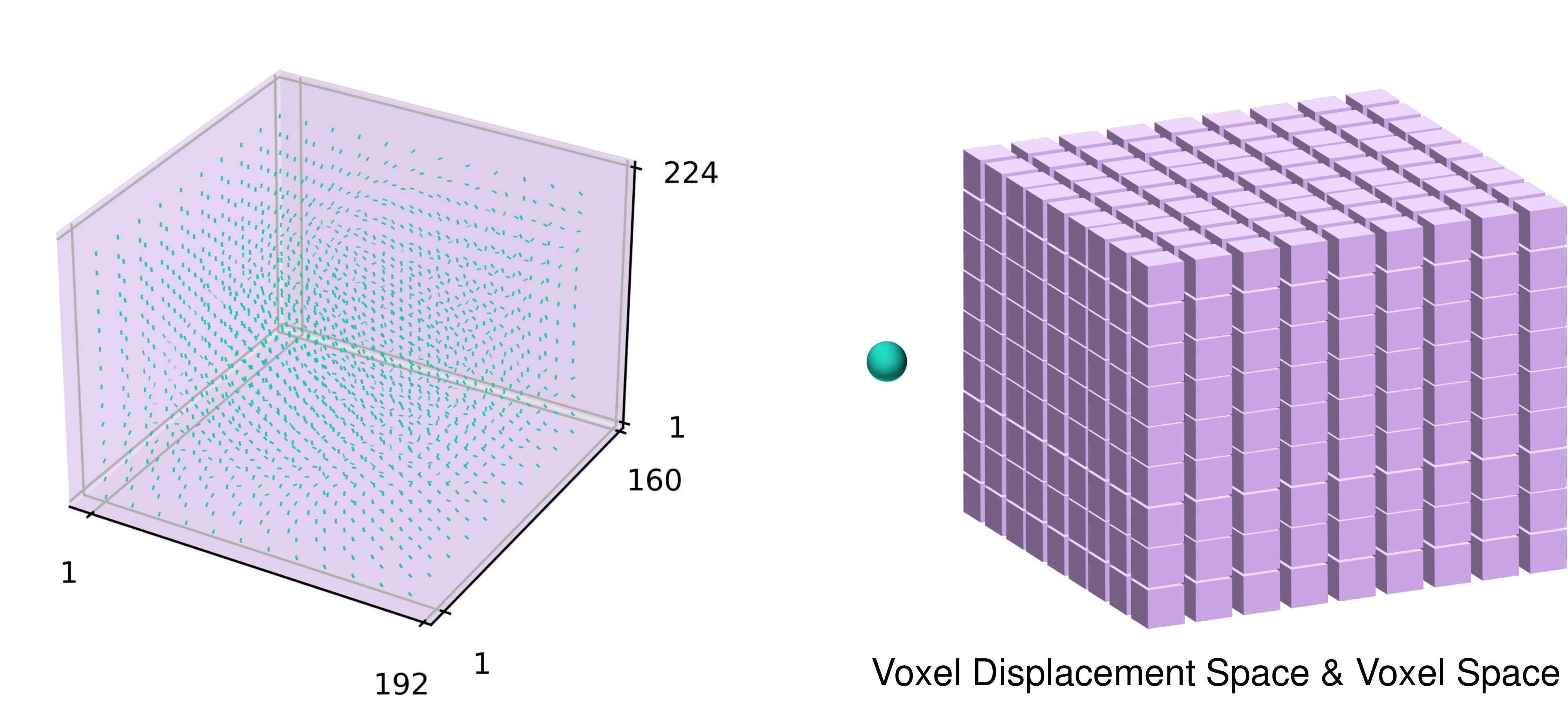}
  \vspace{-10pt}
  \caption{Displacement fields computed from the IXI brain dataset. The left figure plots the displacement vectors in voxel averaged over the whole training data (average lengths of these vectors along $x$-axis, $y$-axis, and $z$-axis are 2.1 voxels, 2.3 voxels, and 1.4 voxels, respectively). The right figure is an illustration of the left figure where vectors are represented by a sphere, the size of which is much smaller than the cubic which represents the true size of the volumetric image.}
\label{fig:voxel_distribution}
\vspace{-10pt}
\end{figure}

The motivation for this work can be found in Fig. \ref{fig:voxel_distribution}, where we plotted the average voxel displacement fields of the IXI brain dataset estimated by TransMorph. We notice that the average length of displacements along $x$-axis, $y$-axis, and $z$-axis are 2.1 voxels, 2.3 voxels, and 1.4 voxels, respectively. These displacements are very small compared to the actual volumetric size (160$\times$192$\times$224) of the image. Therefore, we argue that it may not be necessary to adopt a transformer to model long-range dependencies for deformable image registration. Instead, we propose a large kernel U-Net (LKU-Net) by increasing the effective receptive field of a vanilla U-Net with large kernel blocks and show that our LKU-Net outperforms TransMorph by using only 1.12\% of its parameters and 10.8\% of its mult-adds operations.

\section{Related Works}
\noindent {\bf{U-Net-based registration}}: First published in 2015, U-Net \cite{ronneberger2015u} and its variants have proved their efficacy in many image analysis tasks. VoxelMorph  \cite{balakrishnan2018unsupervised,balakrishnan2019voxelmorph}, one of the pioneering works for medical image registration, used a five-layer U-Net followed by three convolutional layers at the end. The network receives a stacked image pair of moving and fixed images and outputs their displacements. To train the network, an unsupervised loss function was used which includes a warping layer, a data term, and a regularization term. VoxelMorph has achieved comparable accuracy to state-of-the-art traditional methods (such as ANTs \cite{avants_ANTS}) while operating orders of magnitude faster. Inspired by its success, many subsequent registration pipelines \cite{zhang2018inverse,Zhao_2019_ICCV,Mok_2020_CVPR,jia2021learning} used the U-Net style architecture as their registration network backbone. Among them, some works \cite{Zhao_2019_ICCV,jia2021learning} cascaded multiple U-Nets to estimate deformations. An initial coarse deformation was first predicted and the resulting coarse deformation then refined by subsequent networks. This coarse-to-fine method usually improves the final performance, but the number of cascaded U-Nets is restricted by the GPU memory available, so training them on large-scale datasets may not be feasible with small GPUs. In this work we show that, without any cascading, a single U-Net style architecture can already outperform transformer-based networks. 

{\bf{Transformer-based registration}}: Transformer \cite{NIPS2017_3f5ee243} is based on the attention mechanism, and was originally proposed for machine translation tasks. Recently, this architecture has been rapidly explored in computer vision tasks \cite{dosovitskiy2020image,liu2021swin}, because it successfully alleviates the inductive biases of convolutions and is capable of capturing extreme long-range dependencies. Some recent registration methods \cite{chen2021vit,chen2021transmorph,zhang2021learning} have embedded the transformer as a block in the U-Net architecture to predict deformations. Building on a 5-layer U-Net, Zhang et al. \cite{zhang2021learning} proposed a dual transformer network (DTN) for diffeomorphic image registration, but such a dual setting requires lots of GPU memory and greatly increases the computational complexity. As such, DTN can only include one transformer block at the bottom of the U-Net. Chen et al. \cite{chen2021vit} proposed ViT-V-Net by adopting the vision transformer (ViT) \cite{dosovitskiy2020image} block in a V-Net style convolutional network \cite{milletari2016v}. To reduce the computational cost, they input encoded image features to ViT-V-Net instead of image pairs. Their results were comparable with VoxelMorph. The authors of ViT-V-Net \cite{chen2021vit} later improved upon ViT-V-Net and proposed TransMorph \cite{chen2021transmorph} by adopting a more advanced transformer architecture (Swin-Transformer \cite{liu2021swin}) as its backbone. They conducted thorough experiments and showed the superiority of their TransMorph \cite{chen2021transmorph} over several state-of-the-art methods. In this work, we show that our proposed LKU-Net can outperform TransMorph on both inter-subject and atlas-subject brain registration tasks. We conclude in the end that fully convolutional U-Net architectures are still competitive in medical image registration.

\begin{figure}[t!]
  \centering
  \includegraphics[width=0.8\linewidth]{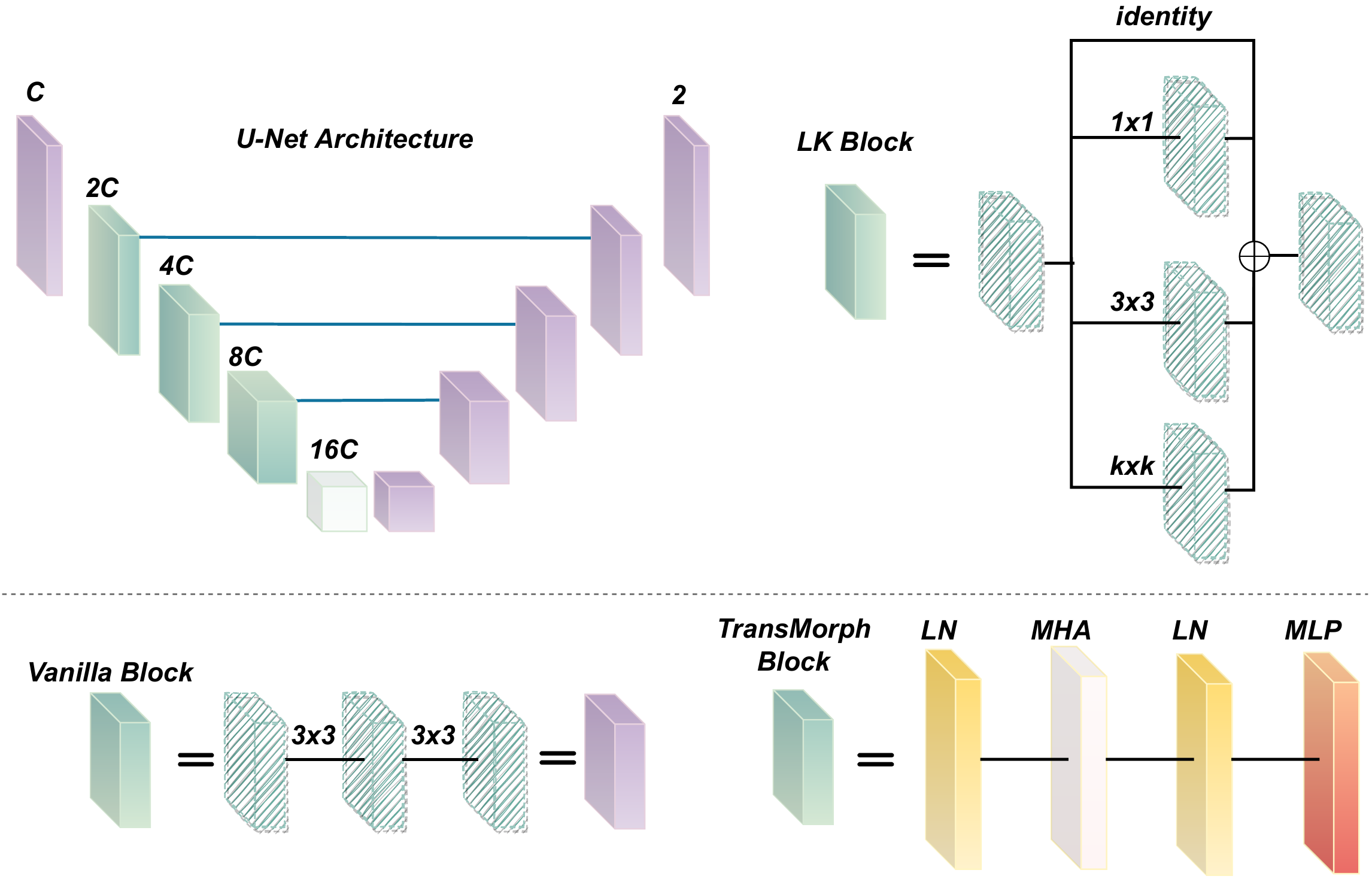}
  \vspace{-10pt}
  \caption{Blocks used in the vanilla U-Net, TransMorph, and LKU-Net. The vanilla U-Net uses the same blocks in both the encoder and decoder, each of which consists of multiple sequential 3$\times$3 convolutional layers. TransMorph replaces four convolutional blocks in the encoder with transformer-based blocks, each of which is built on a combination of layer norm (LN), multi-head self-attention (MHA), and multi-layer perceptron (MLP). For a fair comparison with TransMorph, we use four LK blocks in LKU-Net, each of which contains one identity shortcut and three parallel convolutional layers (that have the kernel sizes of 1$\times$1, 3$\times$3, and $k\times k$, respectively). The outputs of each LK block are then fused by an element-wise addition.}
\label{fig:illustration}
\end{figure}

\section{Large Kernel U-Net (LKU-Net)}
\noindent {\bf{Large kernel (LK) block}}: According to \cite{araujo2019computing}, it is easy to compute that the receptive field of a vanilla 5-layer U-Net is large enough to cover the area which could impact the deformation field around a given voxel. However, as per RepVGG \cite{Ding_2021_CVPR} and RepLK-ResNet \cite{ding2022scaling}, in practice the effective receptive field (ERF) of a convolutional network is much smaller than the one we compute. We therefore adopt a LK block to increase the effective receptive field. Specifically, in each LK block, there are four parallel sub-layers, including a LK convolutional layer ($k\times k \times k$), a 3$\times$3$\times$3 layer, a 1$\times$1$\times$1 layer, and an identity shortcut. The subsequent outputs of these sub-layers are then element-wisely added to produce the output of a LK block, as shown in Fig. \ref{fig:illustration}. The parallel paths in each LK block not only handle distant spatial information but also capture and fuse spatial information at a finer scale.

Directly enlarging the kernel size of a convolutional layer leads to the number of parameters growing exponentially. For example, the number of parameters increases by $463\%$ and $1270\%$ when enlarging a 3$\times$3$\times$3 kernel to 5$\times$5$\times$5 and to 7$\times$7$\times$7, respectively. The resulting network is then cumbersome and prone to collapsing or over-fitting during training. The benefits of using the proposed LK block is that both the identity shortcut and the 1$\times$1$\times$1 convolutional layer help the training. These numerical results are listed in our ablation studies (Table \ref{tab:AblationStudy}).

{\bf{LKU-Net}}: We then propose LKU-Net for registration by integrating the LK blocks into the vanilla U-Net as in Fig. \ref{fig:illustration}. In order to perform a fair comparison with TransMorph, which uses four transformer blocks in the contracting path of its architecture, we only use four LK blocks in the contracting path of the proposed LKU-Net. Note that, the proposed LK block is a plugin block that can be integrated into any convolutional network.

\textbf{Parameterization:} The resulting LKU-Net takes a stacked image pair as input, and outputs the estimated deformation. LKU-Net itself has two sets of architecture specific hyperparameters: the number of kernels and the size of the kernels in each convolutional block. For simplicity, we set the number of kernels in the first layer as $C$; then the number of kernels is doubled after each down-sampling layer in the contraction path and halved after each up-sampling layer in the expansion path; the number of kernels in  the last layer is set to 3 for 3D displacements (2 for 2D displacements). On the other hand, though multiple LK blocks are used within our LKU-Net, we use the same kernel size $k\times k\times k$ for all LK blocks and set all other kernels to be 3$\times$3$\times$3.

\textbf{Diffeomorphism:} Besides directly estimating displacements from LKU-Net, we also proposed a diffeomorphic variant, termed LKU-Net-diff, in which the final output of the LKU-Net is a stationary velocity field $\boldsymbol{v}$. We then use seven scaling and squaring layers to induce diffeomorphisms, i.e. the final deformation $\boldsymbol{\phi} = Exp(\boldsymbol{v})$ as in \cite{ashburner2007fast,dalca2018unsupervised}.

\textbf{Network loss:} We adopt an unsupervised loss which consists of a normalized cross correlation (NCC) data term and a diffusion regularization term (applied to either the displacement or velocity field), which are balanced by a hyperparameter $\lambda$. The overall loss is ${\cal L}({\bf{\Theta}})$ is
$
\min_{\bf{\Theta}}  -\frac{1}{N}  \sum_{i=1}^N {\rm{NCC}}( I_1^i \circ  {\boldsymbol{\phi}_i}({\bf{\Theta}}) - I_0^i)  + \frac{\lambda}{N} \sum_{i=1}^N \|\nabla \boldsymbol{v}_i({\bf{\Theta}}) \|_2^2  
$
Here $N$ is the number of training pairs, $I_0$ denotes the fixed image, $I_1$ represents the moving image, ${\bf{\Theta}}$ are the network parameters to be learned, $\circ$ is the warping operator, and $\nabla$ is the first order gradient implemented using the finite differences.

\section{Experimental Results}
\noindent \textbf{Datasets}: We used two datasets in our experiments. First, \textbf{OASIS} dataset \cite{marcus2007open} consists of 416 cross-sectional T1-weighted MRI scans. We used the pre-processed OASIS dataset (including 414 3D scans and 414 2D images) provided by the Learn2Reg 2021 challenge (Task 3) \cite{hoopes2021hypermorph} for inter-subject brain registration. Each MRI brain scan has been skull stripped, aligned, normalized and has a resolution of 160$\times$192$\times$224. Label masks of 35 anatomical structures were used to evaluate registration performance using metrics such as Dice Score. In this dataset, there are 394 scans (unpaired) for training, and 19 image pairs (20 scans) for validation and public leaderboard ranking. We report our 3D results on their validation set in Tabel \ref{tab:oasis}. For fast evaluation of different methods and parameters, in Table \ref{tab:AblationStudy}, we used 414 2D images with size 160$\times$192, each being one slice of its respective 3D volume. We randomly split the data into: 200 images for training, 14 image pairs for validation, and 200 image pairs for testing.

Second, \textbf{IXI} dataset\footnote{IXI data is available in \url{https://brain-development.org/ixi-dataset/}} contains nearly 600 3D MRI scans from healthy subjects, collected at three different hospitals. We used the pre-processed IXI data provided by \cite{chen2021transmorph}. Specifically, we used 576 T1–weighted brain MRI images to perform atlas-to-subject brain registration, in which 403, 58, and 115 images were used for training, validation, and testing, respectively. The atlas is generated by \cite{kim2021cyclemorph}. All volumes were cropped to size of 160$\times$192$\times$224. Label maps of 29 anatomical structures were used to evaluate registration performances by Dice.

\textbf{Implementation details}: The vanilla 5-layer U-Net architecture used in this work was first proposed by \cite{zhang2018inverse} and then used in \cite{Mok_2020_CVPR}, the only change we made was setting all kernels to 3$\times$3$\times$3. 2D U-Net shares the same architecture except that all kernels are 3$\times$3. In all experiments, we used the Adam optimizer, with batch size being set to 1, and the learning rate being kept fixed at $1\times 10^{-4}$ throughout training. Note that for the 3D OASIS registration (Learn2Reg Task 3), following \cite{chen2021transmorph}, we additionally adopt a Dice Loss.

\begin{wraptable}[13]{r}{0.5\textwidth}
\vspace{-30pt}
\setlength\tabcolsep{5pt}
\caption{Ablation and Parameter Studies}
\centering
\resizebox{0.49\textwidth}{!}{
\begin{tabular}{ccccccc}\hline
Method         & Model    & $k$ & $C$ & Identity & 1x1 & Dice        \\\hline
A1          & U-Net    & -       & 8             & -        & -   & 76.16(4.08) \\
A2          & U-Net    & 5       & 8             & -        & -   & 76.25(4.04) \\
A3          & U-Net    & 7       & 8             & -        & -   & 76.41(4.13) \\
A4          & U-Net    & 9       & 8             & -        & -   & 76.33(3.98) \\
A5          & U-Net    & 11      & 8             & -        & -   & 75.80(4.03) \\ \hline
B1          & LKU-Net & 3       & 8             & Y        & N   & 76.26(4.18) \\
B2          & LKU-Net & 3       & 8             & N        & Y   & 76.40(4.06) \\
B3          & LKU-Net & 3       & 8             & Y        & Y   & 76.47(3.98) \\
B4          & LKU-Net & 5       & 8             & Y        & N   & 76.36(4.08) \\
B5          & LKU-Net & 5       & 8             & N        & Y   & 76.30(4.06) \\
B6          & LKU-Net & 5       & 8             & Y        & Y   & 76.51(4.10) \\\hline
C1          & LKU-Net & 7       & 8             & Y        & Y   & 76.55(4.06) \\
C2          & LKU-Net & 9       & 8             & Y        & Y   & 76.45(4.03) \\
C3          & LKU-Net & 11      & 8             & Y        & Y   & 76.31(4.05) \\\hline
D1          & LKU-Net & 5       & 16            & Y        & Y   & 77.19(3.86) \\
D2          & LKU-Net & 5       & 32            & Y        & Y   & 77.38(3.89) \\
D3          & LKU-Net & 7       & 32            & Y        & Y   & 77.52(3.90) \\ \hline
\end{tabular}}
\label{tab:AblationStudy}
\end{wraptable}

\textbf{Ablation and parameter studies}: In Table \ref{tab:AblationStudy}, we compare the registration performance of our LKU-Net with the vanilla U-Net using Dice on 2D OASIS data. Methods A1-A5 in Table \ref{tab:AblationStudy} indicate that using different kernels in the vanilla U-Net affects the network's performance. Specifically, replacing all 3$\times$3 kernels with 5$\times$5 and 7$\times$7 ones improves Dice by 0.09 and 0.25, respectively. However, when using 9$\times$9 and 11$\times$11 kernels, the performance begins to decline. Comparing the results of Methods A1 \& B3, A2 \& B6, A3 \& C1, A4 \& C2, and A5 \& C3, it is easy to see that our LKU-Net outperforms the U-Net when we use the same kernels size $k$, and that LKU-Net is consistently better than the vanilla U-Net (A1).

Meanwhile, the results from B1-B6 suggest that using either the identity shortcut or the 1$\times$1 layer improves the registration performance, and that combining both leads to the best performance. Comparing B3, B6 and C1, we see that the performance of LKU-Net improves when we increase the kernel size. Lastly, comparing D1, D2 and D3, we find that using larger models also improves the performance, i.e., when we increase $C$ from 8 to 16 and to 32 in LKU-Net$_{8,5}$, Dice improves from 76.51 to 77.19 and to 77.38, respectively.

\begin{table*}[t!]
\setlength\tabcolsep{7pt}
\centering
\caption{Performance comparison between different methods on IXI. Note that the listed results (except the last six rows) are directly taken from TransMorph \cite{chen2021transmorph}.}
\vspace{-5pt}
\resizebox{0.99\textwidth}{!}{
\begin{tabular}{ccccccc}
\hline
Model            & & Dice        & & \% of |J|\textless{}=0     & Parameters & Mult-Adds (G)\\\hline
Affine           & & 0.386$\pm$0.195 & & -                      &-&-\\
SyN              & & 0.639$\pm$0.151 & & \textless{}0.0001      &-&-\\
NiftyReg         & & 0.640$\pm$0.166 & & \textless{}0.0001      &-&-\\
LDDMM            & & 0.675$\pm$0.135 & & \textless{}0.0001      &-&-\\
deedsBCV         & & 0.733$\pm$0.126 & & 0.147$\pm$0.050        &-&-\\\hline
VoxelMorph-1     & & 0.723$\pm$0.130 & & 1.590$\pm$0.339        &274,387&304.05\\
VoxelMorph-2     & & 0.726$\pm$0.123 & & 1.522$\pm$0.336        &301,411&398.81\\
VoxelMorph-diff  & & 0.577$\pm$0.165 & & \textless{}0.0001      &307,878&89.67\\
CycleMorph       & & 0.730$\pm$0.124 & & 1.719$\pm$0.382        &361,299&49.42\\
MIDIR            & & 0.736$\pm$0.129 & & \textless{}0.0001      &266,387&47.05\\\hline
ViT-V-Net        & & 0.728$\pm$0.124 & & 1.609$\pm$0.319        &9,815,431&10.60\\
CoTr             & & 0.721$\pm$0.128 & & 1.858$\pm$0.314        &38,684,995&1461.61\\
PVT              & & 0.729$\pm$0.135 & & 1.292$\pm$0.342        &58,749,007&193.61\\
nnFormer         & & 0.740$\pm$0.134 & & 1.595$\pm$0.358        &34,415,851&686.77\\
TransMorph       & & 0.746$\pm$0.128 & & 1.579$\pm$0.328        &46,771,251&657.64\\
TransMorph-Bayes & & 0.746$\pm$0.123 & & 1.560$\pm$0.333        &21,205,491&657.69\\
TransMorph-bspl  & & 0.752$\pm$0.128 & & \textless{}0.0001      &46,806,307&425.95\\
TransMorph-diff  & & 0.599$\pm$0.156 & & \textless{}0.0001      &46,557,414&252.61\\\hline
U-Net$_4$        & & 0.727$\pm$0.126 & & 1.524$\pm$0.353        &279,086&  58.73\\
U-Net-diff$_4$    & & 0.744$\pm$0.123 & & \textless{}0.0001      &279,086&  58.73\\
LKU-Net$_{4,5}$      & & 0.752$\pm$0.131 & & 0.023$\pm$0.018      &522,302&71.00\\
LKU-Net-diff$_{4,5}$     & & 0.746$\pm$0.133 & & \textless{}0.0001  &522,302&71.00\\ 
LKU-Net$_{8,5}$     & & 0.757$\pm$0.128 & & 0.117$\pm$0.058      &2,086,342&272.09\\
LKU-Net-diff$_{8,5}$     & & 0.753$\pm$0.132 & & \textless{}0.0001 &2,086,342&272.09\\\hline
\end{tabular}}
\label{tab:ixi}
\end{table*}

\begin{figure*}[h!]
  \centering
  \includegraphics[width=0.99\linewidth]{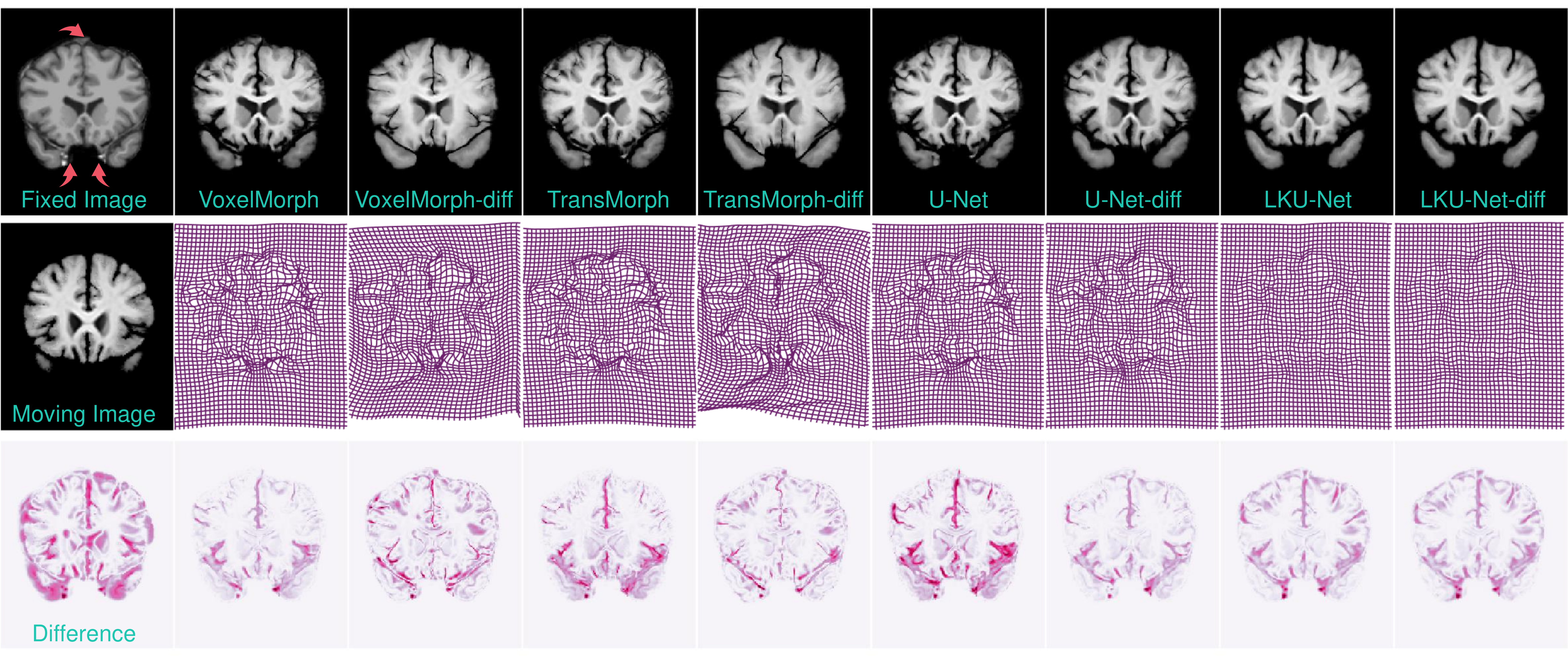}
  \vspace{-10pt}
  \caption{Visual comparison between different methods. The first column displays the fixed image, the moving image and the difference between them. Excluding the first column, from top to bottom we show the warped moving images, the deformations, and the difference maps (between warped moving images and fixed images) of different methods. LKU-Net is able to produce smooth deformations and hence the most realistic warped moving images (see the regions marked with pink arrows).}
\label{fig:visual}
\vspace{-10pt}
\end{figure*}

\textbf{Atlas-to-subject registration on IXI}: To guarantee a fair comparison with TransMorph\footnote{\url{https://github.com/junyuchen245/TransMorph_Transformer_for_Medical_Image_Registration}} on this dataset, we used the exact same data pre-processing, training/validation split, and testing protocol. We report our results in Table \ref{tab:ixi}, where $C$ and $k$ in U-Net$_C$ and LKU-Net$_{C,k}$ respectively denote the number of kernels in the first layer and the kernel size in the LK blocks. 

In the case of non-diffeomorphic registration, U-Net$_4$ achieves 0.727 Dice score which is comparable with VoxelMorph-1 (0.723), VoxelMorph-2 (0.726), ViT-V-Net (0.728), CoTr (0.721), and PVT (0.729), but lower than those of nnFormer (0.740), TransMorph (0.746), and TransMorph-Bayes (0.746). However, LKU-Net$_{4,5}$ outperforms all other competing methods with a Dice score of 0.752. Note that the parameters and mult-adds of LKU-Net$_{4,5}$ are only 1.12\% and 10.8\% of those of TransMorph, respectively. Furthermore, by increasing the number channels from
from LKU-Net$_{4,5}$ to LKU-Net$_{8,5}$, a better Dice score of 0.757 can be achieved.  The visual comparison of estimated deformations is given in Fig. \ref{fig:visual}. For diffeomorphic registration, U-Net-diff$_4$, LKU-Net-diff$_{4,5}$, and LKU-Net-diff$_{8,5}$ achieve Dice score of 0.744, 0.746, and 0.753, respectively. They all outperform TransMorph-diff (0.599) by a large margin and are comparable to TransMorph-bspl (0.752).

\begin{wraptable}[6]{r}{0.5\textwidth}
\setlength\tabcolsep{5pt}
\vspace{-30pt}
\caption{Results of different methods on the Learn2Reg 2021 challenge Task 3.}
\vspace{-5pt}
\centering
\resizebox{0.5\textwidth}{!}{
\begin{tabular}{lcccc}\\\hline
Methods                   & & Dice  &SDlogJ &HdDist95        \\\hline
nnU-Net                   & & 0.8464 ± 0.0159  &0.0668& 1.5003 \\
LapIRN                    & & 0.8610 ± 0.0148 & 0.0721& 1.5139 \\
TransMorph-LC             & & 0.8691 ± 0.0145 &0.0945  & 1.3969 \\
                          \hline
LKU-Net                  & & \textbf{0.8861 ± 0.0150} &0.5169 &\textbf{1.2617}\\\hline
\end{tabular}}
\label{tab:oasis}
\end{wraptable}
\textbf{Subject-to-subject registration on OASIS}: We further evaluated the proposed LKU-Net on the validation set of MICCAI Learn2Reg 2021 challenge (Task 3), and the results are shown in Table \ref{tab:oasis}. Even though our LKU-Net contains only one single network, it has already outperformed TransMorph-LC (which is a large cascaded TransMorph). LKU-Net also outperforms LapIRN with a large margin (2.51\%), which was the winner of Learn2Reg 2021 challenge. As of the submission of this work, our proposed LKU-Net holds first place on the validation leaderboard\footnote{\url{https://learn2reg.grand-challenge.org/evaluation/task-3-validation/leaderboard/}}.


\section{Conclusion}
\noindent With two public 3D brain datasets, we have shown that the proposed method built on a U-Net architecture can outperform modern transformer-based methods for both inter-subject and atlas-to-subject registration tasks. Our LKU-Net is conceptually simple, easy to implement, and has achieved state-of-the-art results, which proves U-Net is still competitive if devised properly. LKU-Net, however, is tested for uni-modality deformable image registration, and we will extend it to estimate deformations across different modalities.
\section{Acknowledgement}
\noindent The research was performed using the Baskerville Tier 2 HPC service. Baskerville was funded by the EPSRC and UKRI (EP/T022221/1 and EP/W032244/1) and is operated by Advanced Research Computing at the University of Birmingham. Xi Jia is partially supported by the Chinese Scholarship Council. Joseph Bartlett is supported by the Melbourne Research Scholarship.
%
%
%
\bibliographystyle{splncs04}
\bibliography{mybibliography}

\begin{thebibliography}{10}
\providecommand{\url}[1]{\texttt{#1}}
\providecommand{\urlprefix}{URL }
\providecommand{\doi}[1]{https://doi.org/#1}

\bibitem{araujo2019computing}
Araujo, A., Norris, W., Sim, J.: Computing receptive fields of convolutional
  neural networks. Distill  \textbf{4}(11), ~e21 (2019)

\bibitem{ashburner2007fast}
Ashburner, J.: A fast diffeomorphic image registration algorithm. Neuroimage
  \textbf{38}(1),  95--113 (2007)

\bibitem{avants_ANTS}
Avants, B.B., Tustison, N.J., Song, G., Cook, P.A., Klein, A., Gee, J.C.: A
  reproducible evaluation of ants similarity metric performance in brain image
  registration. Neuroimage  \textbf{54}(3),  2033--2044 (2011)

\bibitem{balakrishnan2018unsupervised}
Balakrishnan, G., Zhao, A., Sabuncu, M.R., Guttag, J., Dalca, A.V.: An
  unsupervised learning model for deformable medical image registration. In:
  Proceedings of the IEEE Conference on Computer Vision and Pattern Recognition
  (CVPR). pp. 9252--9260 (2018)

\bibitem{balakrishnan2019voxelmorph}
Balakrishnan, G., Zhao, A., Sabuncu, M.R., Guttag, J., Dalca, A.V.: Voxelmorph:
  a learning framework for deformable medical image registration. IEEE
  Transactions on Medical Imaging  \textbf{38}(8),  1788--1800 (2019)

\bibitem{chen2021transmorph}
Chen, J., Frey, E.C., He, Y., Segars, W.P., Li, Y., Du, Y.: Transmorph:
  Transformer for unsupervised medical image registration. arXiv preprint
  arXiv:2111.10480  (2021)

\bibitem{chen2021vit}
Chen, J., He, Y., Frey, E.C., Li, Y., Du, Y.: Vit-v-net: Vision transformer for
  unsupervised volumetric medical image registration. arXiv preprint
  arXiv:2104.06468  (2021)

\bibitem{dalca2018unsupervised}
Dalca, A.V., Balakrishnan, G., Guttag, J., Sabuncu, M.R.: Unsupervised learning
  for fast probabilistic diffeomorphic registration. In: International
  Conference on Medical Image Computing and Computer-Assisted Intervention. pp.
  729--738. Springer (2018)

\bibitem{Ding_2021_CVPR}
Ding, X., Zhang, X., Ma, N., Han, J., Ding, G., Sun, J.: Repvgg: Making
  vgg-style convnets great again. In: Proceedings of the IEEE/CVF Conference on
  Computer Vision and Pattern Recognition (CVPR). pp. 13733--13742 (June 2021)

\bibitem{ding2022scaling}
Ding, X., Zhang, X., Zhou, Y., Han, J., Ding, G., Sun, J.: Scaling up your
  kernels to 31x31: Revisiting large kernel design in cnns. arXiv preprint
  arXiv:2203.06717  (2022)

\bibitem{dosovitskiy2020image}
Dosovitskiy, A., Beyer, L., Kolesnikov, A., Weissenborn, D., Zhai, X.,
  Unterthiner, T., Dehghani, M., Minderer, M., Heigold, G., Gelly, S., et~al.:
  An image is worth 16x16 words: Transformers for image recognition at scale.
  arXiv preprint arXiv:2010.11929  (2020)

\bibitem{hoopes2021hypermorph}
Hoopes, A., Hoffmann, M., Fischl, B., Guttag, J., Dalca, A.V.: Hypermorph:
  Amortized hyperparameter learning for image registration. In: International
  Conference on Information Processing in Medical Imaging. pp. 3--17. Springer
  (2021)

\bibitem{jia2021learning}
Jia, X., Thorley, A., Chen, W., Qiu, H., Shen, L., Styles, I.B., Chang, H.J.,
  Leonardis, A., De~Marvao, A., O’Regan, D.P., et~al.: Learning a
  model-driven variational network for deformable image registration. IEEE
  Transactions on Medical Imaging  \textbf{41}(1),  199--212 (2021)

\bibitem{kim2021cyclemorph}
Kim, B., Kim, D.H., Park, S.H., Kim, J., Lee, J.G., Ye, J.C.: Cyclemorph: cycle
  consistent unsupervised deformable image registration. Medical Image Analysis
   \textbf{71},  102036 (2021)

\bibitem{liu2021swin}
Liu, Z., Lin, Y., Cao, Y., Hu, H., Wei, Y., Zhang, Z., Lin, S., Guo, B.: Swin
  transformer: Hierarchical vision transformer using shifted windows. In:
  Proceedings of the IEEE/CVF International Conference on Computer Vision. pp.
  10012--10022 (2021)

\bibitem{marcus2007open}
Marcus, D.S., Wang, T.H., Parker, J., Csernansky, J.G., Morris, J.C., Buckner,
  R.L.: Open access series of imaging studies (oasis): cross-sectional mri data
  in young, middle aged, nondemented, and demented older adults. Journal of
  cognitive neuroscience  \textbf{19}(9),  1498--1507 (2007)

\bibitem{milletari2016v}
Milletari, F., Navab, N., Ahmadi, S.A.: V-net: Fully convolutional neural
  networks for volumetric medical image segmentation. In: 2016 fourth
  international conference on 3D vision (3DV). pp. 565--571. IEEE (2016)

\bibitem{Mok_2020_CVPR}
Mok, T.C., Chung, A.C.: Fast symmetric diffeomorphic image registration with
  convolutional neural networks. In: IEEE/CVF Conference on Computer Vision and
  Pattern Recognition (CVPR) (June 2020)

\bibitem{qiu2021learning}
Qiu, H., Qin, C., Schuh, A., Hammernik, K., Rueckert, D.: Learning
  diffeomorphic and modality-invariant registration using b-splines. In:
  Medical Imaging with Deep Learning (2021)

\bibitem{ronneberger2015u}
Ronneberger, O., Fischer, P., Brox, T.: U-net: Convolutional networks for
  biomedical image segmentation. In: International Conference on Medical Image
  Computing and Computer-Assisted Intervention. pp. 234--241. Springer (2015)

\bibitem{rueckert1999nonrigid}
Rueckert, D., Sonoda, L.I., Hayes, C., Hill, D.L., Leach, M.O., Hawkes, D.J.:
  Nonrigid registration using free-form deformations: application to breast mr
  images. IEEE Transactions on Medical Imaging  \textbf{18}(8),  712--721
  (1999)

\bibitem{thorley2021nesterov}
Thorley, A., Jia, X., Chang, H.J., Liu, B., Bunting, K., Stoll, V., de~Marvao,
  A., O’Regan, D.P., Gkoutos, G., Kotecha, D., et~al.: Nesterov accelerated
  admm for fast diffeomorphic image registration. In: International Conference
  on Medical Image Computing and Computer-Assisted Intervention. pp. 150--160.
  Springer (2021)

\bibitem{NIPS2017_3f5ee243}
Vaswani, A., Shazeer, N., Parmar, N., Uszkoreit, J., Jones, L., Gomez, A.N.,
  Kaiser, L.u., Polosukhin, I.: Attention is all you need. In: Guyon, I.,
  Luxburg, U.V., Bengio, S., Wallach, H., Fergus, R., Vishwanathan, S.,
  Garnett, R. (eds.) Advances in Neural Information Processing Systems.
  vol.~30. Curran Associates, Inc. (2017)

\bibitem{vercauteren2009diffeomorphic}
Vercauteren, T., Pennec, X., Perchant, A., Ayache, N.: Diffeomorphic demons:
  Efficient non-parametric image registration. NeuroImage  \textbf{45}(1),
  S61--S72 (2009)

\bibitem{zhang2018inverse}
Zhang, J.: Inverse-consistent deep networks for unsupervised deformable image
  registration. arXiv preprint arXiv:1809.03443  (2018)

\bibitem{zhang2021learning}
Zhang, Y., Pei, Y., Zha, H.: Learning dual transformer network for
  diffeomorphic registration. In: International Conference on Medical Image
  Computing and Computer-Assisted Intervention. pp. 129--138. Springer (2021)

\bibitem{Zhao_2019_ICCV}
Zhao, S., Dong, Y., Chang, E.I.C., Xu, Y.: Recursive cascaded networks for
  unsupervised medical image registration. In: The IEEE International
  Conference on Computer Vision (ICCV) (October 2019)

\end{thebibliography}
\end{document}